\documentclass{article}

\usepackage{amsmath} \usepackage{amsfonts} \usepackage{amssymb}

\usepackage[pdftex]{graphicx} 
\usepackage{color}

\begin{document} 
\begin{sloppypar}

\noindent{\Large Nonspecific biological effects of weak magnetic fields\\depend on molecular rotations}

\medskip\medskip\medskip
\noindent{\bf Vladimir N. Binhi$^{1,2,\#}$} \& {\bf Frank S. Prato$^{3,4}$}

\medskip
\noindent $^{1}$A.M. Prokhorov General Physics Institute, Moscow, Russia\\
$^{2}$M.V. Lomonosov Moscow State University, Moscow, Russia\\
$^{3}$Lawson Health Research Institute, Ontario, Canada\\
$^{4}$University of Western Ontario, Ontario, Canada

\medskip
\noindent
{\footnotesize $^{\#}$vnbin@mail.ru}

\begin{abstract} The radical pair mechanism is a leading hypothesis in animal magnetic navigation. This mechanism associates the magnetic sense with the visual system, the radical pairs in cryptochromes of the eye retina being specialized magnetic receptors that modulate rhodopsin-mediated photoreception. There are also nonspecific magnetic effects in biology, which occur mostly by chance and originate from the interaction of weak magnetic fields with the magnetic moments dispersed all over the organism at the microscopic level. The radical pair mechanism cannot explain this type of response for many reasons. We have previously shown that the above interaction has a finite probability of resulting in an observable. Here, we develop our physical model of nonspecific magnetic effects for the case of magnetic moments located in rotating molecules. We generalize the results of recent experiments on gene expression in plants in a constant magnetic field, and show that the precession of the magnetic moments that reside on rotating molecules can be slowed relative to the immediate biophysical structures. In quantum mechanical language, the crossing of the quantum levels of magnetic moments conjointly with molecular rotations explain nonspecific magnetic effects and leads to magnetic field-dependences that are in good agreement with the experiment. \end{abstract}


\subsection* {Introduction}

A large number of different biological effects can be observed at weak magnetic fields (MFs) in the range 0.1--100~$\mu $T, e.g.\,\cite{Breus.ea.2016}. For  migratory animals that have formed their specific magnetic sense in the course of evolution, the initial transduction mechanism is apparently associated with spin-correlated radical pairs in retina cryptochromes, or with the radical pair mechanism (RPM), e.g.\,\cite{Schulten.ea.1978, Hore.and.Mouritsen.2016}. In contrast, there is a so-called nonspecific response to magnetic fields that is characteristic of all organisms and manifests itself only occasionally, at random effective combinations of electromagnetic and biochemical/physiological conditions \cite{Binhi.and.Prato.2017.PLoS}. Nonspecific response differs from magnetoreception in navigating animals, is of fundamental importance, and has its own biophysical mechanisms that remains largely unknown.

We have recently shown that the effects of  a hypomagnetic field (HMF) occupy a special place among the nonspecific effects due to their higher magnitude and reproducibility and their capability of providing more precise information on the origin of the MF effects. An accurate definition of HMF is given in \cite{Binhi.and.Prato.2017.PLoS}; it reads mainly that both the static $H$ and the rms value of the variable component $h$ should be much less than the geomagnetic field (GMF) $H_{\rm g} \sim 50~\mu $T. Special terms have also been introduced for those micro objects that react with MF at the beginning of a signal transduction path and thus can be referred to as the targets of the MF. These are primary physical targets, i.e., magnetic moments, and molecular or biophysical MF targets, or sensors that carry the moments and can change depending on the state of the moments. The existence of HMF effects can be illustrated in terms of quantum mechanics. The Zeeman sublevels of a magnetic moment that are split in a MF, degenerate, or cross, provided their width, of the order of $\hbar /\tau $, where $\tau $ is the thermal relaxation time and $\hbar $ is the reduced Plank constant, becomes comparable with the Zeeman splitting  $\hbar \gamma H$, where $\gamma $ is the gyromagnetic ratio. Then a critical MF $H \sim 1/\gamma\tau $ defines the MF magnitude below which changes should occur at the quantum level.

A physical mechanism for HMF effects has been proposed \cite{Binhi.2016, Binhi.and.Prato.2017}, which considers the dynamics of non-uniformly precessing magnetic moments in biophysical targets, or MF sensors that are not specialized MF receptors. Magnetic effects occur as the consequence of a significant slowing the precession of magnetic moments, or, in a quantum picture, their quantum levels crossing. Therefore, in what follows, the mechanism is referred to as the ``level crossing'' mechanism (LCM).

Recent experiments show that a weak static MF  affects the expression of some genes in \textit{A. thaliana} \cite{Dhiman.and.Galland.2018t}, where a few well-resolved peaks can be observed in MF-dependences. This work has also shown that a MF reversal produces an asymmetric response. Similar effects on gene expression have been previously reported \cite{Bertea.ea.2015}.

It seems unlikely that the RPM could explain these results due to experimental \cite{Harris.ea.2009, Bertea.ea.2015,  Dhiman.and.Galland.2018t, Hoyto.ea.2017} and theoretical \cite{Binhi.and.Prato.2017.PLoS} reasons. At the same time, similar multi-peak magnetic response in \textit{E. coli} \cite{Binhi.ea.2001} had been well described by the interference mechanism \cite{Binhi.1997.BF} extended to the case of molecular rotations. The interference mechanism and LCM are cognate: both predict preferred angular positions of their objects. For this reason, it was relevant to explain the spectral character of gene expression in \textit{A. thaliana} and asymmetric response at the MF reversal in terms of molecular rotations using the LCM theory.

Rotations are ubiquitous at the level of molecular processes, particularly in many of those related to DNA, RNA, and ATPases. As previously suggested \cite{Binhi.2000, Binhi.ea.2001}, rotations of the molecules that carry the precessing magnetic moment can significantly affect the process in which the magnetic moment initiates subsequent biophysical events. 

Below we extend the LCM to the case where the molecular surrounding of magnetic moments rotates with a natural biological speed. The probability of secondary biophysical events in the MF sensors that reside on rotating molecules is estimated and $H$-dependences of this probability is calculated. Based on the comparison with the multi-peak $H$-dependence \cite{Dhiman.and.Galland.2018t} for gene expression in \textit{A. thaliana} growing in static MF (SMF), a conclusion will be made that rotating macromolecules are involved in nonspecific response to MF in organisms, and the molecular MF sensors reside on such macromolecules.

\subsection* {Rotations}

It is interesting to note, that reproducible and large nonspecific magnetic effects are observed in systems with pronounced processes involving gene expression: neurite outgrowth \cite{Blackman.ea.1993}, cephalic regeneration in planarians \cite{Jenrow.ea.1995}, morphological changes during embryogenesis \cite{Delgado.ea.1982}, response to heat shock \cite{Prato.2015}, cell growth and gene expression \cite{Bertea.ea.2015} in plants, the proliferation of neuroblastoma cells \cite{Mo.ea.2014} and of nerve stem cells \cite{Fu.ea.2016}, gravitropism \cite{Belyavskaya.2004}. The combined action of MF and the X-ray \cite{Politanski.ea.2013} and of MF and heavy ions \cite{Lebedev.ea.2014} can be seen as the interference between DNA repair and nonspecific magnetic effects. There is also strong dependence of magnetic effects on the genetic modification of organisms \cite{Alipov.ea.1996, Celestino.ea.1998, Mo.ea.2012b, Mo.ea.2013, Xu.ea.2017}.  All the above suggests that gene expression could be a prerequisite for MFs to elicit an effect.

Transcription and translation are known to be accompanied by the rotation of  macromolecules. For example, ribosome and its parts rotate in the process of translation \cite{Xie.2015}. Motion of RNA polymerase and helicase along the DNA helix is accompanied by relative rotation of the enzyme and DNA. Several examples of the macromolecular rotations in \textit{E. coli} cells with speeds from a few to a few hundred rps are listed in \cite{Binhi.ea.2001}: DNA topoisomerase, DNA and RNA polymerases, F$_{\textrm{O}}$F$_1$-ATPases, the flagellar motor. Other contenders for rotating macromolecules could be: myosin rotation about an actin filament at $1.5$--$2.5$~rps  \cite{Sase.ea.1997}, rotation of microtubules induced by dynein at $1$--$4$~rps \cite{Vale.ea.1988}.

Galland \cite{Dhiman.and.Galland.2018t} considers chloroplast F$_{\textrm{O}}$F$_1$ ATPases with a concomitant functioning of V-ATPases most likely MF sensors in plants. Rotation of the chloroplast F$_{\textrm{O}}$F$_1$-ATPase motor, through the redox state of plastids, modulates gene expression and depends on light, which is in agreement with the observed light-dependence of magnetic response.

Thus, when modeling magnetic biological effects one must take into account rotations of the immediate environment of magnetic moments, i.e., the rotations of the MF sensor itself.

If an organism is experiencing a spatial shift along any axis in the laboratory frame, then shifted are all its elements such as atoms and molecules together with their magnetic moments. It might seem that if the body rotated, all of its elements would rotate with it also. However, this is not the case with respect to elementary precessing angular momenta and their magnetic moments. Due to physical laws, rotational motion of the MF sensor body does not transmit the torque to the precessing angular momentum. For example, while holding a gyroscope, it is easy to move its body linearly, but it is difficult to roll out or slow down its rotor --- this would require specific movements. Similarly, a magnetic moment precesses mostly independently of the molecular enclosure, although the moment's thermal relaxation is due to the interaction with it.

If the angular velocity $\mathbf \Lambda $ of the rotating body is close to that of the magnetic moment precession $ \gamma {\bf H}$, there will be a situation similar to the temporary slowing of the moment, in the rotating frame of the body, or a level crossing. A HMF effect arises, although the MF is not a HMF. This mechanism resembles a roulette --- the ball falls in the cell when the angular velocities of the ball and of the roulette wheel coincide. Similarly, HMF effect occurs when there is a coincidence of the angular velocities of the precession and of the sensor body.

In general, an organism may contain various types of sensors that rotate with several different speeds. In this case, when scanning the dc MF magnitude, the effect like a HMF effect will occur sequentially for sensors rotating at different speeds. A quantum interference mechanism of nonspecific magnetic effects that considers processes involving rotations has been previously proposed in \cite{Binhi.2000} and developed in \cite{Binhi.2002} p.\,266--275. This mechanism predicted that rotations affect the $H$-dependences. The \textit{a posteriori} prediction was in agreement with experiment \cite{Binhi.ea.2001}, where a complex multi-peak $H$-dependence of a cell culture response to MF has been observed.

Essentially, it has been shown that the HMF effect at $h=H=0$ decreases for rotating MF sensors. It is easy to deduce that the characteristic feature of the HMF effect --- an abrupt change in the measured value with reducing $H$ to zero --- is merely shifted to some other value of $H$ that depends on the target rotation speed. In the $H$-dependence, this generates a kind of ``window,'' where the magnetic effect can be observed. The ion interference mechanism's \cite{Binhi.2000} capability of explaining multi-peak SMF dependences is reproduced in the LCM. At the same time, the LCM allows one to determine the thermal relaxation time of the primary MF target (magnetic moments) from experimental data, which is a significant advantage.

Recently we have shown that the magnitude of a hundred different HMF effects correlate neither with the HMF value, nor with the period of the exposure to HMF, nor with their product, or ``dose'' \cite{Binhi.and.Prato.2017.PLoS}. The lack of correlation strongly suggests that there is no MF sensor with the same magnetic properties for all organisms. The lack of a general MF sensor and the necessity of gene expression for nonspecific effects suggest their random nature and their link to the varying rotations that bring about the dispersal of magnetic properties of the MF sensors.

Thus, $H$-dependences of nonspecific effects in the SMF could be a new form of magnetic spectroscopy capable of measuring the physical characteristics of molecular processes of transcription and replication. Important information about the molecular processes involving rotations can be extracted from the experiment supported by a quantitative theory capable of explaining such dependences.

\subsection* {Mechanism}

In \cite{Binhi.and.Prato.2017} we proposed a  physical mechanism of nonspecific response to MF in organisms, which considers a nonuniform precession and thermal relaxation of a magnetic moment in the MF of parallel dc and ac components --- the LCM. There are no quantum transitions caused by such a MF, and therefore it is sufficient to use the classical model of the Larmor precession. Here we extend the model to the case of rotating molecular structure that encloses the precessing magnetic moment.

The LCM assumes that a biological response occurs where, within the relaxation time period, the MF disturbs the dynamics of the magnetic moment so that the deviation from the state of the undisturbed uniform precession becomes significant. In \cite{Binhi.and.Prato.2017} we derive the following equation of motion in spherical co-ordinates for a precessing magnetic moment under applied dc and ac magnetic fields, \begin{equation} \label{fi} \varphi (t) =\gamma Ht+\frac{\gamma h}{\Omega } \,{\mathrm{sin} \left( \Omega t\right)\ } \end{equation} where $\varphi $ is the precession phase, or an azimuth angle, $\gamma $ is the gyromagnetic ratio, $H$ is the dc MF magnitude, and $h$ and $\Omega $ are the amplitude and frequency of the ac MF. In the absence of the ac MF, at $h=0$, a uniform precession takes place: $\varphi (t)=\gamma Ht$, where $\gamma H$ is the Larmor frequency. For a weak MF effect to cause a biological response, this background precession should be disturbed. As shown, an effective disturbance can be achieved either by a significant decrease in the dc MF or by modulating the dc field with an ac one. In the first case, the precession stops: $\dot{\varphi }=0$.

Further, a concept of ``reaction'' was defined: it is the change in the state of the biophysical environment that immediately surrounds the precessing moment. The probability of the event to occur in the time interval $[t-\tau /2,t+\tau /2]$ was assumed to be e.g.\,\cite{Ross.1996} \begin{equation} \label{pe} p\left(t,\tau \right)=1-{\mathrm{exp\ } \left[-\int^{t+\tau /2}_{t-\tau /2}{\lambda }\left(u\right)\mathrm{d}u\right]\ } \end{equation} where $\lambda $ is the density of the Poisson process and $\tau $ is the thermal relaxation time. If magnetic moments precess, their oscillations transform to those in the rate of downstream events $\lambda $. The simplest idealization of this fact is \begin{equation} \label{la} \lambda =\beta \left[1+{\mathrm{cos} \left(\varphi -\xi \right) }\right] \end{equation} where the proportionality factor $\beta $ with the dimension of frequency is introduced, and $\xi $ denotes the random magnetic moment direction that maximizes $\lambda $. The density is minimal when the directions $\varphi $ and $\xi $ are opposite.

Based on Eqs.~\ref{fi}--\ref{la}, the probability of biophysical events $P(H,h,\Omega,\gamma,\tau,\beta ) \equiv \langle p \rangle _{t, \xi} $ averaged over $t$ and $\xi $ was derived which depends on six quantities. First three of these are MF variables, and three others are MF sensor parameters: the gyromagnetic ratio $\gamma $ and the parameters $\tau $ (relaxation time) and $\beta $ (mean temp of the biophysical events initiated by the precessing moments). Difference $P(H,h,\Omega,...) - P(H,0,\Omega,...)$ describes the probability change under ac/dc MF exposure and shows the maximum effect at $h=1.8 H$ and $\Omega = \gamma H $, which is in agreement with many experiments, see a review in \cite{Binhi.2002} p.\,307--314.

At $h=0$, under a dc MF that is decreasing from the geomagnetic field $H_{\rm g}$ to an HMF $H$, the probability change $\Delta P \equiv P(H,0, ... ) - P(H_{\rm g},0, ... )$ equals \begin{equation} \label{PofH} \Delta P (H,\gamma,\tau,\beta )  \approx - \frac{1}{4}{\beta }^2{\tau }^2     e^{-\beta\tau}  \textrm{sinc} ^2 \left( \gamma H\tau /2\right) \end{equation} where $P(H_{\rm g},0, ... )$ is assumed to be $P(\infty,0, ...)$. This equation explains the HMF effect, that occurs when the dc MF is decreased from $H_{\rm g}$ to HMF $H$, see also Fig.~6 of \cite{Binhi.and.Prato.2017}.

Below, in order not to complicate the model by including all possible types of MF exposure, only this case of a constant MF will be considered. For convenience in what follows, the HMF effect will be associated with $-\Delta P$, i.e., with a positive magnitude. It will be shown that rotations of MF sensors shift the MF at which a biological effect gets an abrupt change from the range of MFs close to zero to higher MFs. We will refer to this as a static magnetic field (SMF) effect. Extension of the theory above explained to a rotating MF sensor requires modification of Eq.~\ref{la}.

First, besides $\beta $ --- the undisturbed rate of biophysical events generation ---  another parameter, the depth of modulation of $\beta $ should be introduced. However, the value of this additional parameter would be determined by a specific biophysical mechanism indicating the type and characteristics of the MF sensor, whereas our model, being a physical one, should be formulated independently of biophysical details. Therefore, we leave this parameter equal to unity and note that the absolute value of the actual magnetic effects can be either larger or smaller than those calculated. Thus, the purpose of the theory below is to calculate the MF-dependencies rather than the magnitude of a magnetic effect. Only the specific form of such dependencies can be compared with experiment.

Second, let ${\bf m} $ be a unit vector in the direction of a precessing magnetic moment and $\bf b$ one in the direction that is associated with the biophysical environment. This vector is defined so that  $\lambda $ acquires a maximum value when $\bf m$ points along $\bf b$. Then, a generalization of Eq.~\ref{la} includes a scalar product of ${\bf m}$ and ${\bf b}$: \begin{equation} \label{lag} \lambda =\beta (1+ {\bf m}{\bf b} ) \end{equation}

\begin{figure} \centering \includegraphics[width=0.36\linewidth]{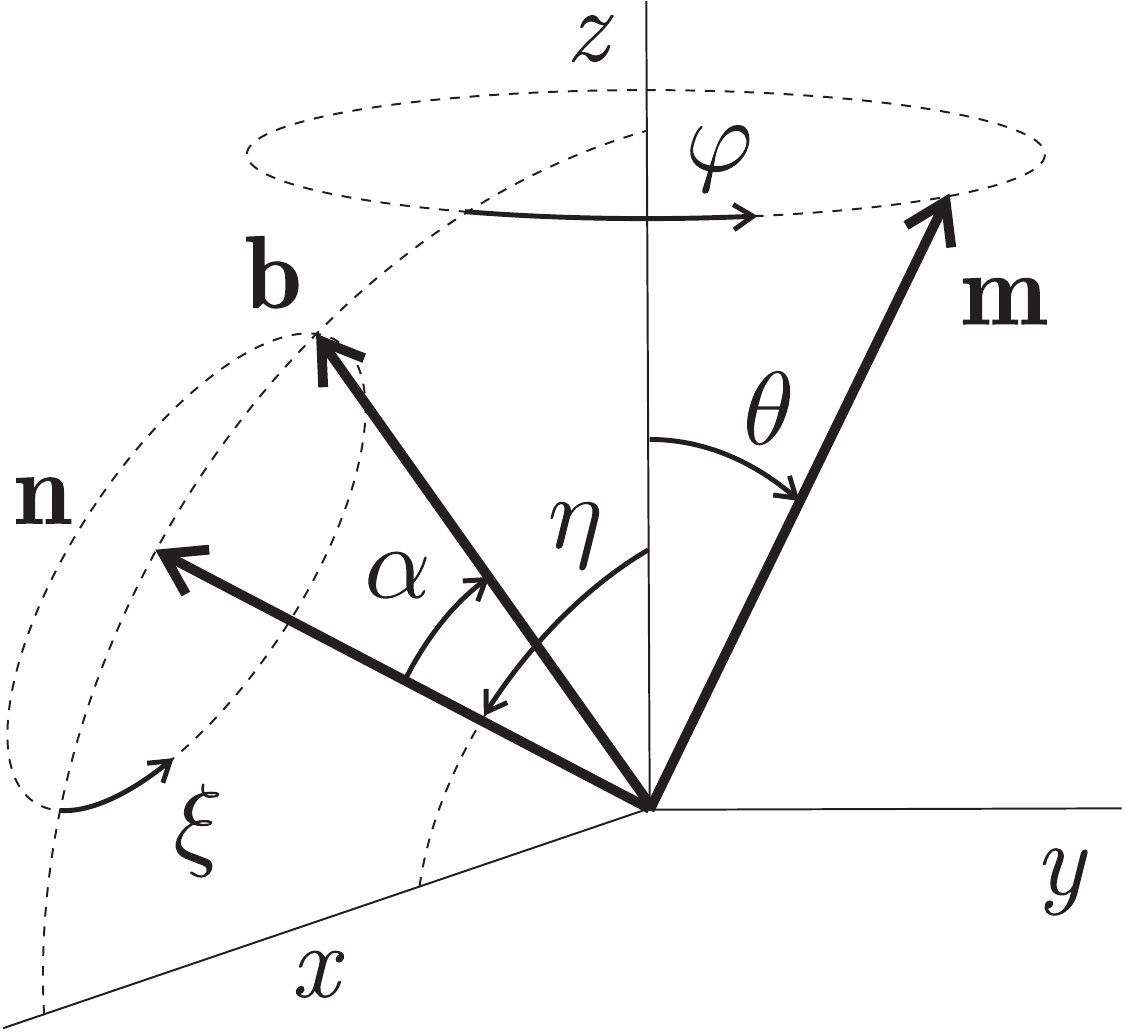} \caption{Unit vectors and their polar angles in a spherical coordinate system. Dash-line circles are the tracing of the precession of vector $\bf m$ and the rotation of vector $\bf b$ around $\bf n$; vector ${\bf b}$ is shown at $t=0$. } \label{fig:2} \end{figure}

Let now ${\bf n} \equiv \mathbf{\Lambda} /\Lambda$ be the unit vector of the MF sensor rotation, and the Cartesian coordinates are chosen so that axis $z$ is directed along $\bf H$ and axis $x$ is in the plane formed by vectors $\bf H$ and ${\bf n} $, Fig.~\ref{fig:2}. Let the target rotate about  $\bf n$ with angular frequency $\Lambda $, which means a rotation of vector $\bf b$ so that $\xi = \Lambda t$. Note that due to subsequent time averaging and practical incommensurability of the rates of precession and rotation, their phases are not significant, and we assume the vectors $\bf m$ and $\bf b$ at $t=0$ be in the $xz$ plane. Note also that for a convenient definition of ${\bf b}$, we use angle $\alpha $ as one between ${\bf b}$ at $t=0$ and $\bf n$ rather than its polar angle. Vector $\bf m$ of the magnetic moment precesses about $\bf H$ so that $\varphi = \gamma H t$, Eq.~\ref{fi}.

With the above notation, one can find the Cartesian components of vectors $\bf m$ and $\bf b$: $m_x =\sin(\theta ) \cos (\gamma H t ) $, $ m_y= \sin(\theta ) \sin (\gamma H t )$, $m_z= \cos(\theta ) $, $b_x = \cos (\alpha) \sin(\eta ) - \sin(\alpha ) \cos (\eta ) \cos (\Lambda t) $, $b_y= - \sin(\alpha ) \sin (\Lambda t) $, $b_z=  \cos(\alpha ) \cos(\eta ) + \sin(\alpha ) \sin(\eta ) \cos (\Lambda t) $. Then the density of biophysical events $\lambda = \lambda (t,\beta,\theta,\eta,\alpha,\gamma,H,\Lambda)$ initiated by precession can be calculated from Eq.~\ref{lag}. Substitution of $\lambda $ into Eq.~\ref{pe} gives a result that needs to be further averaged over time and random variables. One should keep in mind that random variables $\theta $ and $\eta,\alpha $ are of different type with regard to averaging, for the following reasons.

The polar angle $\theta $ is that of the magnetic moment vector $\bf m$ at $t=0$. The orientation of this vector should be considered random for each MF sensor separately. This means the averaging over $\theta $ should be performed unconditionally. The initial orientation of $ \bf m $ is arbitrary in a full solid angle, hence the result generally must be averaged over the azimuth and polar angles. However, the result does not depend on the azimuth angle, and only the unconditional averaging over $ \theta $ remains.

In contrast, the positions of the rotation vector $\bf n$ and target vector ${\bf b} $ at $t = 0$ given by angles $ \eta $ and $ \alpha $ respectively, Fig.~\ref{fig:2}, has a definite value for each specific target. Averaging over these angles makes sense only if they have variable random values in different targets. Note that this is not the case if the targets of the same type have a predominant orientation. For example, many plant cells are oriented in a certain way relative to the gravity vector. Consequently, rotations of macromolecules carrying MF sensors could inherit this preferred orientation in the form of a more or less definite orientation of the vector $\bf n$ and vector $\bf b$ at $t = 0$. Thus, it is relevant to study two cases, (i) deterministic and (ii) stochastic with uniformly distributed random values of $\eta $ and $\alpha $.

Substituting Eq.~\ref{lag} in Eq.~\ref{pe} and performing  averaging over time and $\theta $, one can derive the probability of the secondary events initiated by a precessing magnetic moment, i.e., $P=P(H, \gamma, \tau,\beta, \Lambda, \eta, \alpha )$, \begin{equation*} P=  \frac1{\pi T} \int_0^{\pi} \int_0^T \left[ 1-{\rm exp} \left( -\int^{t+\tau /2}_{t-\tau /2} \lambda \, {\rm d}u \right) \right]  {\rm d}t \,\sin (\theta )   {\rm d} \theta \end{equation*} where $T \equiv 2\pi /|\gamma H - \Lambda |$ is the period of a two-frequency oscillating process. Finally, we arrive at an expression for the probability change that is suitable for comparison with experiment, \begin{equation} \label{Pfin} \Delta P (H, \gamma, \tau,\beta, \Lambda, \eta, \alpha ) \equiv P(H,\gamma, ... ) - P(\infty,\gamma, ... ) \end{equation} As is seen, three new variables are added to $\Delta P$ as compared with the case of fixed MF sensors, see Eq.~\ref{PofH} --- it is the speed $\Lambda $ of the MF sensor rotation and the angles that define axis of rotation $\bf n$ and vector $\bf b$ of the sensor.

As an analytical evaluation of Eq.~\ref{Pfin} would be too cumbersome, a few different cases have been studied numerically. SMF effect has been calculated in different modes at $\beta =2$. 

Dependences on $\gamma H$ at $\eta =0$ and $\alpha = \pi/2$ show: a) a resonance-like response, $\Lambda =0$; b) the asymmetry of the response regarding a MF reversal, $\Lambda =10$ and $15$, $\tau = 1$. Dependences at $\tau = 1$, $\gamma H =10$, and $\Lambda = 10$ on $\alpha$ with $\eta =0$ and on $\eta $ with $\alpha = \pi/2$ show relatively wide ranges of angles $\alpha$ and $\eta $ where magnetic effect exist.   

If $\Lambda =0$ or the vector ${\bf b}$ is parallel to $\bf n$, i.e., $\alpha =0$, this obviously comes to the case without rotation. The half-width of peak at $H=0$ is consistent with Eq.~\ref{PofH}: half-width is defined by the argument of the cardinal sine function where it rapidly changes, i.e., $\Delta H \sim 1/(\gamma \tau)$.

In case of rotation, the most pronounced result occurs where the axis of rotation coincides with the $z$-axis and the MF sensor vector $\bf b$ is perpendicular to the $z$-axis, that is $\eta =0$ and $\alpha = \pi / 2$. Then, the probability of biophysical events that are caused by precessing magnetic moments has a resonance-like peak provided the angular velocity of rotation is in certain relation with the MF vector: $\mathbf {\Lambda } \sim  -\gamma  {\bf H}$.  The position of the peak shifts proportionate to $\Lambda $. This means that the level crossing, or a slow precession, occurs at values $H \sim \Lambda /\gamma $. Note that the effect is not symmetric with respect to the MF reversal, which can be directly tested in experiment. 

The $\gamma H$-position of the peak of magnetic responce is independent of the values of all variables other than $\Lambda $. This enables one to study the $\eta $- and $\alpha $-dependences that are mentioned above. While the effect is in its maximum, one can examine how it depends on the orientation of the axis of rotation $\bf n$ and that of the sensor vector $\bf b$. These dependences are rather smooth. This leaves a chance to observe the SMF effect even for arbitrary or unknown values of these angles.

As was said above, the rotation axes of macromolecules can inherit a preferred orientation of cells; then the values of $\eta$ and $\alpha $ can be considered definite. How could the result change, where these values are random rather than definite? Let both angles, as polar ones, be distributed in the range $[0, \pi $). Since the position of the peak in $\Delta P (\gamma H)$ does not depend on these variables, it is sufficient to average the effect magnitude over these angles only in the peak.

The result shows that the peak height decreases by more than an order of magnitude, from $7${\%} to about $0.5${\%}. The latter is an order of magnitude larger than the RPM effect observed in the GMF-like MFs. However, this value is still small for reliable explanation of the SMF biological effects. This means that for the SMF effect to occur, some kind of rotation ordering is desirable. However it is not necessary for non-rotating MF sensors that show their $14$\,{\%}  HMF effect (as follows from Eq.~\ref{PofH} at $\tau\beta=2$) independently of molecular rotations.

\subsection* {Discussion}

Although there are more than two hundred articles, documenting HMF effects in organisms \cite{Binhi.and.Prato.2017.PLoS}, these data have not, in general, investigated the MF-dependence needed for comparison to the predictions of the LCM or LCMr --- its ``rotation'' extension. However, recently, a set of MF stimulus-response curves for gene expression has been obtained in a study of plant germination in MFs ranging from about $0.5$ to $188$~$\mu $T \cite{Dhiman.and.Galland.2018t}, making it possible to compare theory and experiment. The experimental data demonstrate at least two well-resolved peaks. In order to fit these data, we have assumed that the MF sensors of the same type rotate at two different speeds, thus forming two peaks in the $H$-dependence. The sum of their equal contributions has been calculated from Eq.~\ref{Pfin} with the values of variables that provide maximum effect and assuming magnetic moments of the same type precess inside biophysical structure/structures rotating at two  speeds, $46$ and $116$ $\mu $T/$\gamma $.  

The result shows a good ageement with experimental $H$-dependence of the relative transcript amount of \textit{rbcl} (large subunit of ribulose bisphosphate carboxylase/oxygenase) in seedlings of \textit{A. thaliana} raised for $120$~h under broad-band blue light. Experimental data were kindly provided by P.\,Galland; article \cite{Dhiman.and.Galland.2018t} contains information on the  methods, magnetic exposure, and conditions for this and similar experiments with other genes and strains. 

Two general features of gene expression that have been observed experimentally in germinating plants are essential for a theoretical discourse. This is 1) the presence of resonance-like peaks in $H$-dependences of the SMF effect, and 2) the absence of a symmetry in the response with regard to SMF reversal \cite{Dhiman.and.Galland.2018t}. These features are observed in the $H$-dependences of the expression of a few genes in a few \textit{A. thaliana} strains. As our results demonstrate, the LCM modified to molecular rotations is able to describe, if not to explain, these key features.

Based on the above $H$-dependence and the LCMr predictions, one might speculate on what is the primary MF target.

As follows from the above formulae, two relations should be satisfied in order for this theory to be consistent with the experiment: \begin{equation}  \label{test} \gamma H_{\textrm{p}} \sim \Lambda, ~~ \gamma \tau \Delta H \sim 1 \end{equation} where $H_{\textrm{p}}$ is the location and $\Delta H$ is the half-width of a peak in the $H$-dependence of the probability of biophysical events initiated by the precessing magnetic moment at the sensor rotation with angular speed $\Lambda $. As follows from  the comparison of experiment with theory, exemplary experimental values for $H_{\textrm{p}}$ and $\Delta H$ are about $0.5$ and $0.1$~G, or $50$ and $10$~$\mu $T, respectively. Could this connection of theory with experiment reveal characteristics or the nature of the primary physical target? To do this, one can test Eqs. \ref{test}, while substituting $\gamma $ and calculating $\tau $ and $\Lambda $. The choice is spinning or orbiting electrons, protons or magnetic isotopes, bound ions, and molecular gyroscopes \cite{Binhi.ea.2002.PRE}.

For an electron spin magnetic moment ($\gamma = 1.76\times 10^7$~rad\,G$^{-1}$s$^{-1}$), thermal relaxation time, as follows from Eqs. \ref{test}, is a few times greater than $100$~ns --- a maximum expected for relevant electrons in a wet tissue under physiological temperature. However, this would require the MF sensor to rotate at about $1.5\times 10^6$~rps, which is not realistic.

A proton ($\gamma = 2.68\times 10^4$~rad\,G$^{-1}$s$^{-1}$) and proton-like spin magnetic moments would require rotation speeds of about $2\times 10^3$~rps, which is more realistic, and relaxation time of about $0.5$~ms, which is acceptable for the spin of bound protons. However, a proton's influence on the immediate environment is expected to be weaker than that of electrons, as protons are less mobile and carry much smaller magnetic moment.

The involvement of orbiting bound ions, e.g., Ca$^{2+}$ ($\gamma = 241$~rad\,G$^{-1}$s$^{-1}$), as a primary physical target is not realistic. Although in this case Eqs. \ref{test} give $\Lambda \sim 20$~rps, $\tau $ should be too large, about $30$~ms, which is impossible due to the fact that this time is mostly picoseconds in the order of magnitude.

Finally, a big molecular rotor, like a GLU residue ($\gamma \approx 70$~rad\,G$^{-1}$s$^{-1}$), would require the speed of rotation of about $6$~rps and the thermal relaxation time $0.1$~s. This is more likely, although not without difficulty, as a relatively large cavity of $1.5$-nm radius and free of water molecules is needed to house the rotating group inside the folding protein \cite{Binhi.2002, Binhi.ea.2002.PRE}.

Evidently, $H$-dependencies are not yet sufficient to identify the nature of the primary targets. Probably, the theory should take into account the often intermittent character of molecular rotations, like in RNA and ATPase. In addition, $\Omega $-dependencies obtained in the same organism under the ac/dc MF exposure, as explained in \cite{Binhi.and.Prato.2017.PLoS}, could provide direct information on the gyromagnetic ratio of the primary targets.

LCMr is a general mechanism that explains nonspecific response to MF regardless of the biophysical construct that hosts a precessing magnetic moment. Other mechanism that take into account the biophysical medium is molecular gyroscope mechanism \cite{Binhi.ea.2002.PRE} that could explain the above discussed effects \cite{Dhiman.and.Galland.2018t}. Future experiments are needed to discriminate between two molecular rotor mechanisms --- LCMr and the gyroscopic one --- regarding their involvement in formation of the multi-peak $H$-dependence of the nonspecific response.

Studies on the HMF effects are important for future space flights that are featured by MFs more than thousand-fold smaller than the GMF. For this reason, when studying the magnetic effects on Earth, researchers model the space conditions by correcting for gravity. This is usually achieved with clinostats that rotate samples so that the gravity vector in the frame of the sample is averaged to zero. Due to the influence of rotations on the nonspecific magnetic effects, the widely accepted interpretation of the results obtained in clinostats should be revised.

In summary: (i) molecular rotations are a significant factor affecting nonspecific magnetic effects in organisms, (ii) the Level Crossing Mechanism as applied to rotations, or LCMr, that we have introduced here explains key features of the observed MF-dependences including resonance-like peaks and asymmetry with regard to the SMF reversal, (iii) further insight into the nature of nonspecific response to MF can be provided by multi-peak $H$-dependences of biological effects under controlled outer rotations and, if in plants, under exposure to HMF/SMF of different orientation with respect to the gravity vector, and (iv) fundamental biological information on the molecular rotations can be obtained from the shift of the peaks. The non-invasive extraction of such fundamental information on the rotation of sub-cellular structures introduces the potential to use LCMr as a new biological spectroscopy.

\end{sloppypar} 
\end{document}